\documentclass[prl,twocolumn,showpacs,showkeys]{revtex4}
\usepackage{slashbox}
\usepackage{amsmath,graphicx,amsfonts,bm,amssymb}

\begin{document}

\title{Measurement based Controlled Not gate for topological qubits\\ in a Majorana fermion  quantum-dot hybrid system}

\author{Zheng-Yuan Xue}

\affiliation{Laboratory of Quantum Information Technology, and
School of Physics and Telecommunication Engineering, South China
Normal University, Guangzhou 510006, China}

\date{\today}

\begin{abstract}
We propose a scheme to implement Controlled Not-gate for topological qubits
in  a quantum-dot-Majorana-fermion hybrid system. Quantum
information is encoded on pairs of Majorana fermions, which live on
the the interface between topologically trivial and nontrivial
sections of a quantum nanowire deposited on an s-wave
superconductor. A measurement-based two-qubit Controlled-Not gate
is produced with the help of parity measurements assisted by the
quantum-dot and followed by prescribed single-qubit gates. The
parity measurement, on the quantum-dot and a topological qubit, is
achieved by the Aharonov-Casher effect.
\end{abstract}



\maketitle


Recently, physical implementation of quantum computers has attracted
much attention. One of the main difficulties of scalable quantum
computation is  decoherence of quantum information.  A promising
strategy against decohenrence is based on the topological idea
\cite{tqc} where gate operations depend only on global features of
the control process, and thus largely insensitive to local noises.
Topological ordered states emerge as a new kind of states of quantum
matter beyond the description of conventional Landau¡¯s theory
\cite{wen}. A paradigmatic system for the existence of anyons is a
kind of so-called fractional quantum Hall states \cite{wen}.
Alternatively, artificial spin lattice models are also promising for
observing these exotic excitations, e.g., Kitaev models
\cite{kitaevspin} are most famous for demonstrating anyonic
statistics \cite{han,lucy,xuefs} and braiding operations for
topological quantum computation.

For universal quantum computation, one needs non-Abelian anyon to
serve as qubit.  With the potential applications in topological
quantum computation, Majorana fermions (MF) with non-Abelian
statistics have attracted strong renewed interests. MF are a kind of
self-conjugate quasi-particles induced from a vortex excitation in
$p_x+ip_y$ superconductor \cite{tqc}. However, due to the
instability of the $p$-wave superconducting states, its
implementation remains an experimental challenge.
Recently, it is recognized that topologically protected states may be most easily
engineered in 1D semiconducting nanowires deposited on an
\emph{s}-wave superconductor \cite{1d1,1d2,1d3,jiang}, which
provides the first realistic experimental setting for Kitaev's 1D
topological superconducting state \cite{kitaev}. Note that the quench
dynamics of this model across a quantum critical point is also presented \cite{md}.

Although not universal for quantum computation, parity protected
quantum logical operations schemes with a pair of MF as a
topological qubit are proposed
\cite{saupra,qc1,qc2,xue,qc3,qc4,qc5}, which mainly focus on
braiding operations of single qubit \cite{saupra,qc1,qc2,xue} and
quantum information tranfer between a topological qubit and a
quantum bus \cite{qc3,qc4,qc5}. As the absence of topologically
ordered system  that is universal for quantum computation and
extremely difficult to perform certain topology changing operations,
present research of universal quantum computation with MF, as well
as other exotic topological qubits, need to supplement braiding
operations with topologically unprotected operations. These
operations can be error-corrected for a high error-rate threshold of
approximately 0.14 \cite{threshold}. However, such a high error
threshold may still prove difficult using unprotected operations
within a topological system.

One of the main difficulties in implementing topological quantum computation
lies in the difficulty of braiding of MF from different topological qubits for entangling operation.
Therefore, a topological quantum bus, usually in a topological and conventional qubit hybrid
architecture, would be of great help, where error rates below 0.14
have already been achieved \cite{tbus}. A topological quantum bus
makes the implementation much easier as the quantum bus can transfer the
quantum information between different topological qubits, and thus only
braiding within a topological qubit, to realize single-qubit operation, is needed for quantum computation.
Here, we propose a scheme to implement Controlled Not-gate for topological qubits
in a quantum-dot-MF hybrid system.  The architecture consists of 1D
semiconducting wires deposited on an \emph{s}-wave superconductor,
under certain conditions, the endpoints of such wires support
localized zero-energy MF. Qubit is encoded on a pairs of MF, the realization of parity
measurements on a QD and a topological qubit, together
with control and measurement on the QD state and single-qubit gates
on the target topological qubit, is able to generate a two-qubit
Controlled-Not (CNOT) gate \cite{loss}. In this sense, this scheme
is a measurement-based scenario. Supplemented with arbitrary single qubit
rotation, which is already proposed in such MF quanutm dot hybrid system \cite{1d3,qc1},
universal quantum computation can be realized.


\begin{figure}[tbp]
\centering \includegraphics[width=8cm]{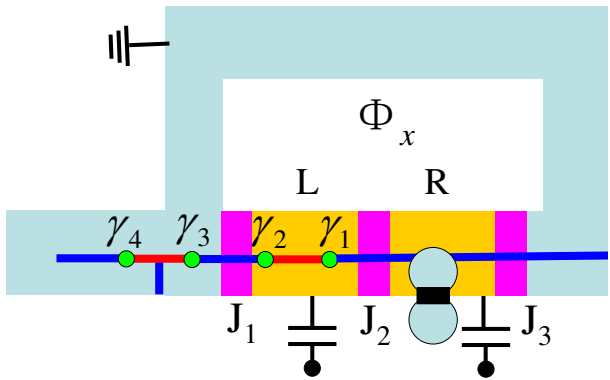} \caption{(Color
online) Superconducting flux qubit with three Josephson junctions
(pink) and an enclosing magnetic flux $\Phi_x$. MF (green dots) are
induced at the interface between a topologically trivial (blue) and
a topologically nontrivial (red) section of an quantum nanowire. The
right superconducting island is located by a semiconductor double
quantum-dot qubit. Gate electrodes (not shown) can be used to move
the MF along the wire.} \label{fig1}
\end{figure}

The setup we consider is shown in Fig. \ref{fig1}, which is  a
spin-orbit coupled semiconducting wire deposited on an \emph{s}-wave
superconductor, which, together with Josephson junctions, form a
superconducting flux qubit configuration. Applying a magnetic field
perpendicular to the superconductor surface, the Hamiltonian
describing such a wire is \cite{1d1}
\begin{eqnarray}
H&=& \int \bigg{[} \psi^\dagger_x
\bigg(-\frac{\hbar^2\partial_x^2}{2m}-\mu - i \hbar u {\hat e}\cdot
{\sigma}\partial_x +V_B \sigma_z\bigg)\psi_x
  \nonumber \\
  && + (|\Delta| e^{i\varphi} \psi_{\downarrow x} \psi_{\uparrow x} +
  h.c.)\bigg{]}dx,
  \label{h}
\end{eqnarray}
where $\psi_{\alpha x}$ corresponds to electrons with spin $\alpha$,
effective mass $m$, and chemical potential $\mu$; the third term
denotes spin-orbit coupling with $u$ the strength, and ${\sigma} =
(\sigma_x,\sigma_y,\sigma_z)$ is the vector of Pauli matrices; the
fourth term represents the energy shift due to the magnetic field;
and the terms in the second line are the spin-singlet pairing from
the $s$-wave superconductor via proximity effect. The interplay of
Zeeman effect, spin-orbit coupling, and the proximity to an s-wave
superconductor drive the wire into a chiral $p$-wave superconducting
state \cite{1d1,1d2}, providing that the wire is
long compared to the superconducting coherence length ($\xi\simeq
40$ nm for the superconducting substrate being Nb).  For $|\mu|<\mu_c = \sqrt{V_B^2 -|\Delta|^2}$
the topological phase with end MF emerges, or a topologically
trivial phase. Thus, applying a gate voltage uniformly allows one to
create or remove the MF.  To avoid gap closure, \cite{1d3} a "keyboard" of
local tunable gate electrodes to the wire is used to
control whether a region of the wire is topological or not. For InAs
quantum narowire, assuming $|V_B| \sim 2|\Delta|$ and $\hbar u \sim
0.1$eV{\AA}, the gap for a 0.1$\mu$m wide gate is of order 1K
\cite{1d3}. The QD used here is semiconductor double quantum dot
molecule with one dot deposit on the superconductor of the flux
qubit circuit while the other is not. We assume here that there is a
galvanic isolation between the superconductor and semiconductor, so
that there is no charge transfer between them \cite{qc4}. Remarkably, one can
also realize this QD using InAs nanowires \cite{wire}, which is previously used for supporting
MF, and thus reduce the experimental difficulties of implementation.


A pair of MF can be combined into a complex fermion. The fermion
parity operator  $n_p$ has eigenvalues $-1$ and $+1$ for  states
$|0\rangle$ and $|1\rangle$, respectively. But, the two states
differ by fermion parity, which prevents the coherent superposition of the two. Therefore,
for the purpose of quantum computation, where coherent superposition is inevitable,
we combine  four MF to form a topological qubit. In this way, coherent superposition
is permitted for the two encoded qubit subspaces with same fermion parity.
Without loss of generality, we can use the subspace with
fermion parity is even as the encoded qubit states, i.e., the two states of the topological
qubit are encoded as $|00\rangle$ and $|11\rangle$ of the four MF.


We now turn to the problem of reading out the two qubit states, which
is one of the preliminary requirements for quantum computation purpose. As
the states of the two pairs of MF in a topological qubit are always
the same, detecting one of them can then fulfil the purpose of
distinguish them. As noted above, the two states $|0\rangle$ and $|1\rangle$ are
different in fermion parity, so they can be distinguished by
$n_p$. Without loss of generality, we choose to detect the pair of
$\gamma_1$ and $\gamma_2$ while move $\gamma_3$ and $\gamma_4$  out
of the flux qubit circuit, also our measurement circuit, as shown in Fig. \ref{fig1}. Note that
the topological property of the wire will not be interrupted by the junctions if the Josephson
junctions' thickness is much smaller than the superconducting
coherence length $\xi$. To measure the
parity of $n_p$, we  use the suppression of macroscopic
quantum tunneling by the Aharonov-Casher effect \cite{ac}: a Josephson vortex
encircling a superconducting island picks up a phase
$\phi=\pi q/e$ determined by the total charge $q$ coupled
capacitively to the superconductor, which includes both the charge
on the superconducting island and on a nearby gate electrode.

Following Ref.  \cite{qc1,ac}, we consider a superconducting flux qubit
with three Josephson junctions, as shown in Fig. \ref{fig1}, where
junctions 1 and 3 have the same Josephson coupling energy $E_J$
while that of junction $2$ is $\alpha E_J$.  The charging energy
of the islands is much larger than   $E_J$ so that the considered qubit works in the flux
regime. The gauge-invariant
phase drops of the three junctions $1,2,3$ are related to the
total magnetic flux $\Phi_{x}$ through the flux qubit loop by the constraint
$\phi_{1}+\phi_{2}+ \phi_{3}=2\pi\Phi_x/\Phi_{0}$ with
$\Phi_{0} = h/2e$ being the  flux
quanta. On the condition that the size of the qubit is sufficiently small,
the flux generated by the circular supercurrent along the loop can be neglected.
Then, the enclosed flux of the qubit contour comes solely from the
external magnetic filed. Then, the superconducting energy of the flux qubit
reads
\begin{equation}\label{u}
U = -E_J\left[\cos\phi_1+\cos\phi_3+\alpha\cos\left(2\pi{\Phi_x
\over \Phi_{0}}-\phi_1-\phi_3\right)\right],
\end{equation}
which exists two lowest degenerate states,
the supercurrent of which flows clockwise and counterclockwise, respectively.
These states are usually defined as the superconducting flux qubit states.
The  qubit states  correspond to the potential energy minima in Eq. (\ref{u}),
and thus the tunneling between the two states requires quantum phase slips.

When $\alpha>1$, the two
energy minima are connected by two different tunneling paths, differ by
$2\pi$ in $\phi_1$ and $-2\pi$ in $\phi_3$. The
interference between the two tunneling paths constitutes the
circulation of a Josephson vortex around both superconducting islands $L$ and $R$.
Accoding to Aharonov-Casher effect, the acquired phase  is $\psi=\pi q/e$ with $q=\sum_{i=L,R}
(en_{p}^{i}+q_{i})$ being the total charge on the two islands
$en_{p}^{i}$ and gate capacitors $q_{i}$ to the qubit loop. As we chose junctions 1 and 3
have the same Josephson coupling energy, the two tunneling paths have the
same amplitude. The interference between the two tunneling paths produces an oscillating
tunnel splitting of the two levels of the flux qubit
\begin{equation}
\Delta  = \Delta_{0} \bigl|\cos\left({\psi \over 2}\right) \bigr|,
\end{equation}
where $ \Delta_{0}$ is the tunnel splitting associated with one
path. Therefore, if $q$ is an odd (even) multiple of the electron
charge $e$, the two tunneling paths interfere destructively
(constructive), and thus the tunnel splitting   is minimum (maximal).

As we only need to distinguish maximal from minimal tunnel
splitting, the flux qubit does not need to have a high quality
factor.  In addition, $ \Delta_{0}\simeq 100 \mu{\rm eV}\simeq 1$K
for parameters in typical experiments of flux qubits \cite{ac},
which should be readily observable by microwave absorption. To make
sure the total charge is solely comes from the two superconducting
islands, one would first calibrate the charge on the gate capacitor
to zero, i.e., $q_{i}=0$, by maximizing the tunnel splitting in the
absence of vortices in the island.  The read-out is  nondestructive,
which is necessary for our proposed way of implementation the
two-qubit CNOT gate. Meanwhile, it is insensitive to sub-gap
excitations in the superconductor as they do not change the fermion
parity.


Following Ref.  \cite{qc4}, as shown in Fig \ref{fig1}, $q$ can also have a quantum component:
charge comes from the QD. Logical basis of QD can be defined as
semiconductor charge qubit $|0\rangle=|0\rangle\otimes|1\rangle$ and
$|1\rangle=|1\rangle\otimes|0\rangle$ with the electron occupies the
lower and upper dot, respectively. Therefore, the qubit basis states
correspond to the electron parity on the upper dot enclosed by the
Josephson vortex circulation. To read-out the MF state, one can
simply set the QD to $|0\rangle$ so that its charge do not interrupt
the measurement. For our purpose, we are also interest in the joint
parity of QD and a pair of MF. Indeed, the flux qubit splitting
energy $\Delta$ is the same for combined topological-QD qubit states
with equal joint parity \cite{qc4}. Thus, measurement of the flux
qubit splitting energy is equivalent to a joint parity measurement
to the states of QD and a pair of MF. In  Ref.  \cite{qc4},
quantum state transfer between topological and QD qubits is achieved by standard quantum teleportation:
the proposed parity measurements supplied with Hadmard gates are equivalent to Bell state measurements.
Here, our scheme for CNOT gate is  a measurement based one.


\begin{figure}[tbp]
\centering \includegraphics[width=8.5cm]{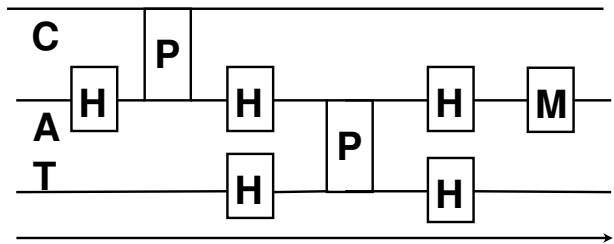}
\caption{Measurement-based CNOT gate for two topological qubits.
Capital letters "A" represents ancillia of QD, "H" is the Hadamard
gate, "C" and "T" represent the control and target qubit,
respectively.  The measurement "M" results of "A"  together with the
outcomes of the two joint parity measurements "P" determine which
operation one has to apply on the "C" and "T" qubit in order to
complete the CNOT gate. The arrowed line in the bottom represents
the sequence of the process. \label{cnot}}
\end{figure}

We next proceed to implement a CNOT  gate between two topological
qubits with the help of QD as an auxiliary. Here we propose a
measurement-based CNOT gate operation \cite{loss}. The relevant
operations are single-qubit rotations, single-QD
rotations/measurements, and effective joint parity measurements for
QD and a topological qubit. The circuit for the CNOT gate is depicted in Fig.
\ref{cnot}. The auxiliary QD is initially prepared  in the  state of
$|0\rangle_A$.   After a Hadamard gate on the QD, the first joint
parity measurement ${P}_1$ in Fig. \ref{cnot} is implemented on the
QD and a pair of MF from "C" qubit.  After Hadamard rotation of the
QD and the target qubit, the second parity measurement ${P}_2$ is
implemented on the QD and a pair of MF in the "T" qubit. Then we
rotate back the QD and the target qubit state by Hadamard gate. The
last step is the measurement of the QD in the $\{|0\rangle,
|1\rangle\}$ basis. The two parity measurement results, together
with the measurement result of the QD determine which single-qubit
gates to be operated on the control and target qubits to generate a
CNOT gate. The relationship between the measurement results and the
gates to be operated is summarized in the table \ref{gate}. After
completing the required gates on the corresponding qubits, it is
straightforward to check that the process is  a CNOT gate operation
between the two qubits.

\begin{table}
\centering \caption{Correspondence between the measurement results
and the gates operated on the control and target qubits. "0" and "1"
represent odd and even parity, respectively.} \label{gate}
\begin{tabular}{  c  c  c  c  c } \hline \hline
"P$_1$" & "P$_2$"  & result of "M" & gate on "C"  & gate on "T"\\
\hline
  1  &  1  &  $|0\rangle_A$  &  I  &  I \\ \hline
  1  & 1  &  $|1\rangle_A$  &  I  & X \\ \hline
 1  &  0  & $|0\rangle_A$  &  Z  &  I \\ \hline
  1  &  0 & $|1\rangle_A$  &  Z  &  X \\ \hline
  0  &  1  &  $|0\rangle_A$ &  I  &  X \\ \hline
  0  &  1  &  $|1\rangle_A$  &  I  & I \\ \hline
  0  &  0  &  $|0\rangle_A$ &  Z &  X \\ \hline
  0  &  0  &  $|1\rangle_A$  &  Z  &  I \\
  \hline \hline
\end{tabular}
\end{table}

Here, we want to emphasize that our implementation is different from that of Ref.
\cite{qc1}, which is alone the line presented in Ref.
\cite{bk}. In the proposal \cite{qc1,bk}, they use the parity measurement  of two MF qubit, to entangle them,
and braiding operations of MF from qubit and ancilla to implement a CNOT gate.
While we use joint MF-QD parity measurements to construct the measurement-based CNOT gate, where only
braiding operations of MF within topological qubits are needed.
In the proposal \cite{qc1,bk}, only one parity measurement  of two MF qubit is needed, thus it is more efficient
in terms of complexity. But, it requires long range braiding operations
between qubit and the ancilla \cite{bk}, which is experimentally challenging.
As for our introducing of a QD qubit to serve as the ancilla, it breaks the
topological protect of the proposal in \cite{bk} in some sense. But, our joint MF-QD parity measurement
is based on the Aharonov-Casher effect, which is also topological in nature and insensitive to local noises.


Before concluding, we want to emphasize that our scheme have the
following distinct merits. (1) As in Ref. \cite{qc4}, one can host
both MF and QD using a single InAs nanowires \cite{wire}, which may
thus reduce the technical challenges of implementation. (2) Here, QD
serves as ancillary qubit.  Therefore, only braiding between MF
within a topological qubit is needed, since braiding of two MF from
different topological qubits will be experimentally challenging. (3)
Measurement of topological qubit uses the Aharonov-Casher effect
\cite{ac}, which is nondestructive. (4) Parity measurement on QD and
a certain topological qubit can be achieved by moving the
topological qubit into the measurement circuit by local tunable gate
on quantum wire \cite{qc1}.

In summary, a
measurement-based two-qubit Controlled-Not gate is produced with the
help of joint parity measurements and followed by prescribed
single-qubit gates.

\bigskip


This work was supported by the NFRPC (No. 2013CB921804),
the NSFC (No. 11004065), the PCSIRT,
and the NSF of Guangdong Province.

\end{document}